\documentclass[pre, tightenlines, superscriptaddress,nofootinbib, twocolumn]{revtex4}
\usepackage{graphicx}
\usepackage{color}
\usepackage{amssymb,amsfonts,amsmath}
\usepackage{natbib}
\usepackage{subfigure}
\usepackage{braket}
\usepackage{lipsum}
\usepackage{amsmath, bm}
\usepackage{hyperref}
\usepackage{epsfig}

\UseRawInputEncoding

\definecolor{orange}{rgb}{1,0.5,0}

\begin{document}

\title{Boundary symmetry breaking of flocking systems}

\author{Leonardo Lenzini} 
\affiliation{Dipartimento di Scienza e Alta Tecnologia, Universit\`{a}
  degli Studi dell'Insubria and Center for Nonlinear and Complex Systems, via Valleggio 11, 22100, Como, Italy}
\affiliation{INFN Sezione di Milano, 20133, Milano, Italy}
\author{Giuseppe Fava} 
\affiliation{Dipartimento di Scienza e Alta Tecnologia, Universit\`{a}
  degli Studi dell'Insubria and Center for Nonlinear and Complex Systems, via Valleggio 11, 22100, Como, Italy}
\affiliation{INFN Sezione di Milano, 20133, Milano, Italy}
\author{Francesco Ginelli} 
\affiliation{Dipartimento di Scienza e Alta Tecnologia, Universit\`{a} degli Studi dell'Insubria and Center for Nonlinear and Complex Systems, via Valleggio 11, 22100, Como, Italy}
\affiliation{INFN Sezione di Milano, 20133, Milano, Italy}

\begin{abstract}
We consider a flocking system confined transversally between two
infinite reflecting parallel walls separated by a distance
$L_\perp$. Infinite or periodic boundary conditions are assumed
longitudinally to the direction of collective motion, defining a ring
geometry typical of experimental realizations with flocking active
colloids. Such a confinement selects a
flocking state with its mean direction aligned parallel to the wall,
thus breaking explicitly the rotational symmetry locally by a boundary
effect.
Finite size scaling analysis and numerical simulations show that
confinement induces an effective mass term ${M_c} \sim L_\perp^{-\zeta}$
(with positive $\zeta$ being the dynamical scaling exponent of the free
theory) suppressing scale free correlations at small
wave-numbers. 
However, due to the finite system size in the transversal
direction, this effect can only be detected for large enough longitudinal
system sizes (i.e. narrow ring geometries). Furthermore, in the longitudinal
direction, density correlations are characterized by an anomalous
effective mass term. 
The effective mass term also enhances the global scalar order parameter
and suppresses fluctuations of the mean flocking direction. These
results suggest an equivalence between transversal confinement and
driving by an homogeneous external field, which breaks the rotational
symmetry at the global level.
\end{abstract}

\maketitle

\section{Introduction}

It is well known that locally aligning self-propelled particles may
achieve global polar order and move collectively in a state known as
{\it flocking} \cite{Ramaswamy, Chate2020}, a collective phenomenon observed in systems as diverse as bird
flocks \cite{Ballerini}, cellular migration \cite{Giavazzi}, motility
assays \cite{Shaller} or active colloids\cite{Bartolo2013}. When polar order
emerges in an isotropic environment, it does so by spontaneously
breaking the underlying continuous rotational symmetry, a fact that
largely determines the large scale physics of flocking systems
\cite{Ginelli2016}. While the non-equilibrium nature of active matter systems
allows flocks to escape the constraints of the
Mermin-Wagner-Hohenberg theorem \cite{MW,H} and achieve long-range order
even in two spatial dimensions \cite{TT2}, the symmetry-broken phase is
nevertheless characterized by massless modes (the {\it
  Nambu-Goldstone modes}) and consequently by long-range
correlations. Notably, long-ranged (or scale-free) correlations have been measured
both in starling flocks \cite{Cavagna2010} and in {\it in vitro}
cellular migration \cite{Giavazzi}: despite the innumerable differences between these
systems and minimal self-propelled particles models, their large-scale
behavior is determined by simple symmetry considerations.
 
The large-scale, long time behavior of flocks is described by the
seminal Toner \& Tu (TT) fluctuating hydrodynamic theory \cite{TT1, TT2,
  TTR}. While the exact value of the TT scaling exponents has long
remained elusive, recent large-scale simulations \cite{Benoit2019} and analytical
breakthrough \cite{Solon} seem to have settled the issue.

Flocking, however, may also take place in anisotropic
environments, where the rotational symmetry is explicitly broken. 
Cell motility, for instance, is known to be sensitive to a wide range of external
gradients of chemical (chemotaxis), mechanical (durotaxis),
and electrical (electrotaxis) origin which often direct cellular
migrations. The response of flocking systems to a global, homogeneous
external field ${\bf h}$ leading the direction of motion of
individual particles has been discussed in the linear regime by
Refs. \cite{Nikos, BFG}. In particular, it is easy to show that the external
field generates an effective mass term and an exponential cut off of
fluctuation correlations. 

In this work, on the other hand, we want to investigate flocking in
the presence of an explicit symmetry breaking at the system
boundaries. This is relevant for many experimental realisations with
active colloids, where confinement by hard boundaries (e.g., within a
ring \cite{Bartolo2013}) is practically unavoidable. Specifically, we
consider a $d$ dimensional flocking system confined in an infinite
channel of width $L_\perp$ by reflecting boundaries conditions in
$d-1$ {\it transversal} spatial directions. Such a confinement
explicitly breaks rotational invariance at the boundaries, selecting
a preferred direction along the free direction  (hereafter denoted as the
{\it longitudinal} direction) and effectively forcing
collective motion along the channel. In finite systems, one can assume
periodic boundary conditions along the free directions, obtaining a
ring geometry as in certain active colloids experimental realisations
\cite{Bartolo}. This setup has been also employed in numerical investigations
of flocking models \cite{TT3, Benoit2019} as a way to suppress the
diffusion of the mean flock orientation in finite systems. The underlying assumption of these
numerical studies is that fluctuations correlations measured in the bulk (that
is, sufficiently far away from  the reflecting boundaries) are left
unperturbed by the boundary symmetry breaking.\\

In this work we verify explicitly these assumptions. While finite size
scaling analysis reveals that transversal confinement induces an
effective mass term -- damping transversal fluctuations and potentially cutting off
correlations above a certain length scale -- it can also be shown
that this exponential cut-off cannot be detected in typical (and
finite) confined geometries, effectively validating the assumption of unperturbed
bulk correlations. \\
Interestingly, this behavior is formally equivalent to that of
a flocking system perturbed (in the linear regime) by an homogeneous external field of
amplitude $h$, provided that
\begin{equation}
h \sim L_\perp^{-z}
\end{equation}
where $z$ is the dynamical scaling exponent of TT theory. This
equivalence extends to other static properties such as the response
and the fluctuations of the global polar order parameter. 


\section{Static correlation functions}
\label{sec1}

\subsection{Brief review of bulk theory}

We first discuss the effect of confinement on the correlation
functions of the relevant slow hydrodynamic fields. In order to do so,
we first briefly review the TT equations \cite{TT2, TTR} which rule the slow, long-wavelength dynamics of the conserved density $\rho({\bf r}, t)$ and velocity ${\bf v}({\bf r}, t)$ fields,
\begin{equation}
\partial_t\rho +\nabla\cdot(\rho{\bf v})=0\;,
\label{rho1}
\end{equation}
\begin{equation} \label{second_v}
    \begin{split}
        \partial_t \textbf{v} &+\lambda_1(\textbf{v} \cdot \nabla)\textbf{v} +\lambda_2 (\nabla \cdot \textbf{v})\textbf{v} + \lambda_3\nabla \lvert \textbf{v} \rvert^2 \\ =& (\alpha - \beta \lvert \textbf{v} \rvert^2 ) {\bf v} -\nabla P_1 -\textbf{v}(\textbf{v} \cdot \nabla P_2) + D_1 \nabla(\nabla \cdot \textbf{v})\\& +D_3 \nabla^2 \textbf{v} +D_2(\textbf{v} \cdot \nabla)^2 \textbf{v} + \textbf{f} \,.
    \end{split}
\end{equation}
Here all the phenomenological convective ($\lambda_i$, with $i=1,2,3$) and viscous ($D_i>0$) coefficients, as well as the two symmetry breaking ones, $\alpha$ and $\beta$, can in principle, depend on $\rho$ and
$|{\bf v}|$ and the pressures $P_{1,2}$ may be expressed as a series
in the density. The additive noise term
${\bf f}$ has zero mean, variance $\Delta$ and is delta correlated in
space and time. For $\alpha >0$, these equations can be linearized around
the homogeneous stationary ordered solution of the fluctuation-less dynamics,
$\rho ({\bf r}) = \rho_0 + \delta \rho ({\bf r})$ and ${\bf v}({\bf r}) =
(p_0 + \delta v_\parallel({\bf r}))\hat{\bf e}_\parallel + {\bf v}_\perp
({\bf r})$, with $\hat{\bf e}_\parallel$ being the unit vector along the
direction of collective motion and $p_0 = \sqrt{\alpha_0/\beta_0}$ \footnote{The pedex ``0''
denotes that the coefficients are evaluated for $\rho=\rho_0$ and
$|{\bf v}|=v_0$.}. Here and in the following the subscripts
$\parallel$ and $\perp$ denote, respectively, the longitudinal and
transversal vector components w.r.t. the reflecting boundaries. Once the fast longitudinal field $\delta v_\parallel({\bf r})$
is enslaved away
\begin{equation}
\delta v_\parallel \approx -\frac{| {\bf v}_\perp|^2}{2p_0}
-\frac{\Gamma_2}{\Gamma_1} \left( \delta \rho - \frac{\partial_t \delta \rho}{p_0 \Gamma_1}
+ \frac{\lambda_4 \partial_{||} \delta \rho}{\Gamma_2}\right) -
\frac{\lambda_2}{\Gamma_1} \nabla_\perp \cdot  {\bf v}_\perp
\label{enslave}
\end{equation} 
(see Ref. \cite{TT3} for more details and the definition of the here
unimportant coefficients $\lambda_4$, $\Gamma_1$ and $\Gamma_2$), 
one is left with two linear equations for the slow
hydrodynamic fields $ \delta \rho ({\bf r})$ and ${\bf
  v}_\perp$ -- the velocity field transversal to $\hat{\bf e}_\parallel$
--  that can be readily solved. \\

Here we focus on the equal time Fourier space fluctuations
correlations (or {\it structure factors}) that for small
wave-numbers $q=|{\bf q}|$ read
\begin{equation}
\label{l_rho}
S_\rho({\bf q}, {\bm \mu}^{(1)}) \equiv \langle |\delta \hat{\rho} ({\bf q})|^2\rangle\sim 
A(\theta_{\bf q}, {\bm \mu}^{(1)})\, q^{-2}
\end{equation}
and 
\begin{equation}
\label{l_v}
S_v({\bf q}, {\bm \mu}^{(1)}) \equiv \langle |\hat{v}_\perp ({\bf q})|^2\rangle\sim 
B(\theta_{\bf q}, {\bm \mu}^{(1)}) \,q^{-2}\;,
\end{equation}
where $\langle \cdot \rangle$ denotes ensemble averages and 
$A(\theta_{\bf q}, {\bm \mu}^{(1)})$ and $B(\theta_{\bf q}, {\bm \mu}^{(1)})$ depend on the
TT equation coefficients, here collectively denoted as ${\bm
  \mu}^{(1)}$, and on the angle
$\theta_{\bf q}$ between ${\bf q}$ and  $\hat{\bf e}_\parallel$.
It is convenient to introduce the longitudinal and transversal components
of the wave-vector, ${\bf q}=({\bf q}_\perp, q_\parallel)$, so that
$q_\parallel = q \cos \theta_{\bf q}$ and $q_\perp \equiv | {\bf q}_\perp |=
 \sin \theta_{\bf q}$.  
The precise expression for $A$ and $B$ is irrelevant for the
following and can be found in Ref. \cite{TTR}, but it should be noted
that the amplitude $A$ is singular in the longitudinal direction,
$A(0, {\bm \mu}^{(1)})=0$, so that the linear behavior of $S_\rho$ for
$q_\perp \to 0$ cannot be determined at this order in TT theory.\\
 
Nonlinearities, moreover, turn out to be relevant in $d<d_c=4$. {\it
  Inter alia}, this is what prevents the linearized fluctuations (\ref{l_v}) to destroy
long-range order due to their non-integrable divergence in $d=2$.
Nonlinearities can be analyzed by a dynamical renormalization group (DRG)
\cite{Tauber} approach. In this procedure, one first averages the slow fields nonlinear equations of motion over
the short-wavelength fluctuations and then, in order to restore the
ultraviolet cut-off of the theory, rescales length-scales, timescales and the slow fields \footnote{One may show that transversal velocity and density fluctuations have the same scaling \cite{TT2}.}
according to $r_\perp=b r_\perp'$, $r_\parallel=b^\xi r_\parallel'$
$t=b^z t'$, $v_\perp=b^\chi v_\perp'$ and $\delta\rho=b^\chi
\delta\rho'$. Here $b>1$ is the (arbitrary) rescaling factor involved
in both the averaging and rescaling steps and the scaling exponents
$\chi$, $\xi$ e $z$ are known as the `roughness', `anisotropy' 
and `dynamical' exponents. By this
procedure one obtains increasingly coarse-grained hydrodynamic
equations with new coefficients ${\mu}^{(b)}$. By a proper choice of
the scaling exponent, the DRG flow typically converges to a nonlinear fixed
point (FP) ${\mu}^*$ that captures the long-wavelength behavior of the theory. It can
be shown that at this nonlinear fixed point 
the structure factors $S_\rho$ and $S_v$ share the same scaling in the
transversal direction, that is for $q_\perp^\xi\gg q_\parallel$,
\begin{equation}
S_v(q_\perp, q_\parallel, {\bm \mu}^*)\sim S_\rho(q_\perp,
q_\parallel,  {\bm \mu}^*) \sim q_\perp^{-\zeta} \,.
\label{nlnSperp}
\end{equation}
where $\zeta \equiv d-1 +\xi +2\chi$. Note that correlations may only
depend on the modulo of the wavenumber in the symmetry unbroken transversal
directions.

In the longitudinal direction, $q_\parallel \gg q_\perp^\xi$, 
transversal velocity correlations scale as
\begin{equation}
S_v(q_\perp, q_\parallel, {\bm \mu}^*)\sim q_\parallel^{-\zeta/\xi} \,.
\label{nlnSpar}
\end{equation}
while the situation is less clear for the density structure factor. 
TT theory predicts 
\begin{equation}
S_\rho(q_\perp, q_\parallel, {\bm \mu}^*)\sim q_\perp^2 q_\parallel^{-2-\zeta/\xi} \,.
\label{nlnSpar2}
\end{equation}
while numerical simulations suggest a different and more complex
behavior\cite{Benoit2019}, which will be briefly discussed in Sec.~\ref{2C}.
This discrepancy, still poorly understood, is probably due to the
singular behavior of the linear density structure factor for
$q_\perp \to 0$ discussed above.\\

The exact value of the nonlinear FP scaling exponent for $d<d_c$ has long remained
elusive due to the pletora of potentially relevant nonlinear terms in
the TT equations. Recent analytical developments, however, suggest
that in the DRG procedure the noise vertex should not acquire graphical corrections due to
the (previously overlooked) gradient structure of the symmetry broken
theory and that a further generalized Galilean
invariance of the theory preserves other nonlinear coefficients at
least for $d=2$. A straightforward calculation leads to the $d=2$
scaling exponents \cite{Solon}
\begin{equation}
\chi = -1/3\;\;\;,\;\;\; \xi=1\;\;\;,\;\;\; z=\zeta=4/3
\label{scalingexp}
\end{equation}
in agreement with large scale simulations results for the Vicsek
model\cite{Benoit2019}. While $\xi=1$ implies, at least in $d=2$, no spatial anisotropy
between the longitudinal and transversal directions, in the following we
derive our result  for the general case $\xi \leq 1$.

\subsection{Scaling of transversally confined flocks}
\label{2B}

In the presence of transversal confinement by reflecting or partially
reflecting walls, the average direction of collective motion aligns
along the channel. The scaling behavior of the structure factors can
then be deduced by a finite size scaling analysis and the request that
for a diverging confinement length $L_\perp \to \infty$ one should
recover the scaling results of bulk Toner \& Tu theory.

Taking into account the dependence on $L_\perp$, the structure factors
$S_\rho$ and $S_v$, here collectively denoted
as $S$, obey the following DRG scaling law
\begin{equation}
S(q_\perp, q_\parallel, {\bm \mu}^{(1)},
L_\perp^{-1})=b^\zeta S(b q_\perp, b^\xi q_\parallel,
{\bm \mu}^{(b)}, bL_\perp^{-1})
\end{equation}
where the scaling of the structure factors has been determined
considering that they are given by the Fourier transform of the equal
time, real space density correlation function. Thus, it involves two
powers of the density fluctuations and one volume element, leading to
the scaling exponent $d-1+\xi+2\chi = \zeta$. 

We choose $bL_\perp^{-1}=1$ which implies $b=L_\perp$. For a
sufficiently large separation, $L_\perp\gg1$, we have ${\bm
  \mu}^{(b)}\simeq {\bm \mu}^*$ and we obtain
\begin{equation}
S(q_\perp,q_\parallel,{\bm \mu}^{(1)},
L_\perp^{-1})=L_\perp^{\zeta}S(L_\perp q_\perp,L_\perp^\xi
q_\parallel,{\bm \mu}^*,1)
\end{equation}
where
\begin{equation}
S(x,y,{\bm \mu}^{*},1) \equiv g(x,y)
\end{equation}
is a universal scaling function. We analyze two different regimes,
depending on whether the behavior of $S$ is dominated by the
longitudinal or transverse wave numbers. 
 
When $q_\perp^\xi\gg q_\parallel$, in the long wavelength limit we
consider the one parameter universal scaling function $w_\perp(x)
\equiv g(x,0) = S(x,0,{\bm \mu}^{*},1)$ which yields the scaling
\begin{equation}
S({\bf q},L_\perp^{-1})=L_\perp^{\zeta}w_\perp(L_\perp q_\perp)
\end{equation}
The behavior of the universal scaling function $w_\perp$ can be
inferred by the request that, for $L_\perp \to \infty$ , the structure factor
scaling coincides with the one of Eq. (\ref{nlnSperp}) and that a tight confinement $L_\perp \to 0$ suppresses the scale free
divergence at small wave-numbers preserving a finite variance, 
\begin{equation}
w_\perp(x)\sim
\begin{cases}
x^{-\zeta} \,\, & x\gg1 \\
\text{constant} \,\,\, & x\ll1
\end{cases}
\end{equation}
The simplest expression for the
scaling function is thus
\begin{equation}
w_\perp(L_\perp q_\perp)\sim\dfrac{1}{(L_\perp q_\perp)^\zeta+G_\perp},
\end{equation}
where $G_\perp$ is a phenomenological parameter depending (among other
things) on microscopic boundary conditions.
Then the structure
factors $S_\rho$ and $S_v$ take the form
\begin{eqnarray}
\label{Sperp}
S_\rho({\bf q},L_\perp^{-1})&\sim&\dfrac{L_\perp^\zeta}{(L_\perp
                              q_\perp)^\zeta+G^{(\rho)}_\perp}=\dfrac{1}{q_\perp^\zeta+G^{(\rho)}_\perp
                              L_\perp^{-\zeta}}\nonumber\\\\
S_v({\bf q},L_\perp^{-1})&\sim&\dfrac{L_\perp^\zeta}{(L_\perp q_\perp)^\zeta+G^{(v)}_\perp}=\dfrac{1}{q_\perp^\zeta+G^{(v)}_\perp L_\perp^{-\zeta}}\nonumber
\end{eqnarray}
for $q_\perp^\xi\gg q_\parallel$ and
with possibly different phenomenological parameters $G^{(\rho)}_\perp $ and
$G^{(v)}_\perp$.

An analogous derivation for the transversal velocity correlation in the
longitudinal case, $q_\perp^\xi\ll q_\parallel$, introduces
the longitudinal scaling function  $w_\parallel(y)
\equiv g(0,y) = S_v(0,y,{\bm \mu}^{*},1)$, whose behavior is also
determined by matching with Eq.  (\ref{nlnSpar}) for $L_\perp \to \infty$,

\begin{equation}
w_\parallel(L_\perp q_\parallel)\sim\dfrac{1}{(L_\perp^\xi q_\parallel)^{\zeta/\xi}+G_\parallel},
\end{equation}
with the phenomenological parameter $G_\parallel$ also depending on
boundary conditions. It follows that when $ q_\parallel \gg q_\perp^\xi$ the structure factor for the transversal velocity takes the form
\begin{equation}
\label{Spar}
S_v({\bf q},L_\perp^{-1})\sim\dfrac{L_\perp^\zeta}{(L_\perp^\xi q_\parallel)^{\zeta/\xi}+G_\parallel}=\dfrac{1}{q_\parallel^{\zeta/\xi}+G_\parallel L_\perp^{-\zeta}}
\end{equation}

Eqs. (\ref{Sperp}) and (\ref{Spar}) express the scaling of the density and
transversal velocity static correlations in Fourier space for
transversally confined flocking. They show how confinement induces an
effective mass term $M_c \sim L_\perp^{-\zeta}$ nominally suppressing the
bulk fluctuations divergence at small wave-numbers. Transversal
confinement, however, implies that $q_\perp$ is a positive
integer multiple of $\pi/L_\perp$. The mass corrections is only
relevant for transversal wave-numbers such that $(L_\perp q_\perp)^\zeta \lesssim
G_\perp$, which implies 
\begin{equation}
G_\perp \gtrsim \pi^\zeta\,.
\label{Cperp}
\end{equation}
We conclude that for $G_\perp$ of order one or smaller the suppression
of the divergence may well not be detectable along the confined
directions. 

This is not the case in the longitudinal direction. For an infinite
system, $q_\parallel$ is unbounded from zero and a plateau in the
transversal velocity structure factor should appear for $q_\parallel \lesssim
G_\parallel^{\xi/\zeta} L_\perp^{-\xi}$. In numerical or experimental
systems one typically deals with periodic boundary
conditions in the longitudinal direction (i.e. a ring geometry) and a
longitudinal size $L_\parallel$ with a corresponding smallest longitudinal
wavenumber  $\pi/L_\parallel$. Boundary effects thus generate a
detectable suppression of the small wavelength divergence for
\begin{equation}
G_\parallel \gtrsim
\left(\frac{L_\perp^\xi}{L_\parallel}\pi\right)^{\zeta/\xi}\,.
\label{Cpar}
\end{equation}

The lack of a clear analytical understanding of the free theory behavior of the
density structure factor in the longitudinal direction prevents us from
formulating a precise scaling form for confined systems, but
nevertheless we still expect that the explicit symmetry breaking will
result in 
\begin{equation}
S_\rho(0, q_\parallel, L_\perp^{-1}) \xrightarrow{q_\parallel \to 0}
\frac{1}{\Sigma(L_\perp)}
\end{equation} 
with $\Sigma(L_\perp)$ an anomalous mass term scaling with a negative power of $L_\perp$.

We have derived the scaling of velocity correlations in the transversal and longitudinal
directions, even if, as previously discussed, recent results predict no
scaling anisotropy ($\xi=1$) in Vicsek flocks. However, this scaling
isotropy does not imply that also the prefactors have to be equal,
leading to the general scaling form
\begin{equation}
\label{Siso}
S_v({\bf q}, L_\perp^{-1})\sim\dfrac{1}{q^{\zeta}+G(\theta_{\bf q}) L_\perp^{-\zeta}}
\end{equation}
where $G(\theta_{\bf q})$ is phenomenological parameter
with $G(0)=G_\parallel$ and $G(\pi/2)=G^{(v)}_\perp$.

Note finally that these results have been derived without any explicit
modelling of the interaction between the active particles and the
confining walls. The only requirement is that the rotational
invariance of the self-propulsion orientation is explicitly broken at
the boundaries. This clearly applies to reflecting or partially
reflecting boundaries, or more generally to any systems in which the
proximity with the wall induces -- directly or indirectly - torques on
the self-propulsion orientation, thus avoiding the trapping of active particles by the wall.

\subsection{Numerical evidence}
\label{2C}

These predictions can be verified considering the Vicsek model (VM)
\cite{Vicsek1995, Chate1} -- the prototypical microscopic
flocking model -- transversally confined by reflecting
boundary conditions. In two spatial dimensions, the discrete time
dynamics of $N$ active particles with position ${\bf r}_i^t$ and
unit self-propulsion orientation ${\bf s}_i^t=(\cos \theta_i^t, \sin \theta_i^t)$  is given by
\begin{equation}
\label{VM}
\theta_i^{t+1}=\text{Arg}\left(\sum_{j\sim i}\vec{s_j}^t\right)+\eta\xi_i^t,\quad\vec{r_i}^{t+1}=\vec{r_i}^t+v_0\vec{s_i}^{t+1},
\end{equation}
where $j \sim i$ indicates that the corresponding sum is carried over
all the particles $j$ within unit distance of $i$ (including $i$
itself) and $\text{Arg}(\vec{u})$ returns the angle defining the
orientation of $\vec{u}$. The active particles move with constant
speed $v_0$ and their orientation is subject to a zero-average and
delta-correlated scalar noise term of amplitude $\eta$, with the $\xi_i^t$
being independent variables uniformly drawn from the interval
$[-\pi,\pi]$. We consider periodic boundary conditions in the
longitudinal direction of size $L_\parallel$ and implement the
transversal reflecting boundaries by the following collision rule
\cite{Fava2024}
\begin{equation}
s_\perp \!\!\to\! - s_\perp\;\;\;,\;\;\; r_\perp
\!\!\to\! 2 B \!- r_\perp
\end{equation} 
with either $B\!=\!0$ (left boundary) or
$B\!=\!L_\perp$ (right boundary), which is applied whenever the
Vicsek dynamics (\ref{VM}) would result in a particle position 
with transversal component $r_\perp$, outside the region $r_\perp
\!\in\! [0,L_\perp]$. \\

\begin{figure}[t!]
\centering
\includegraphics[width=0.48\textwidth]{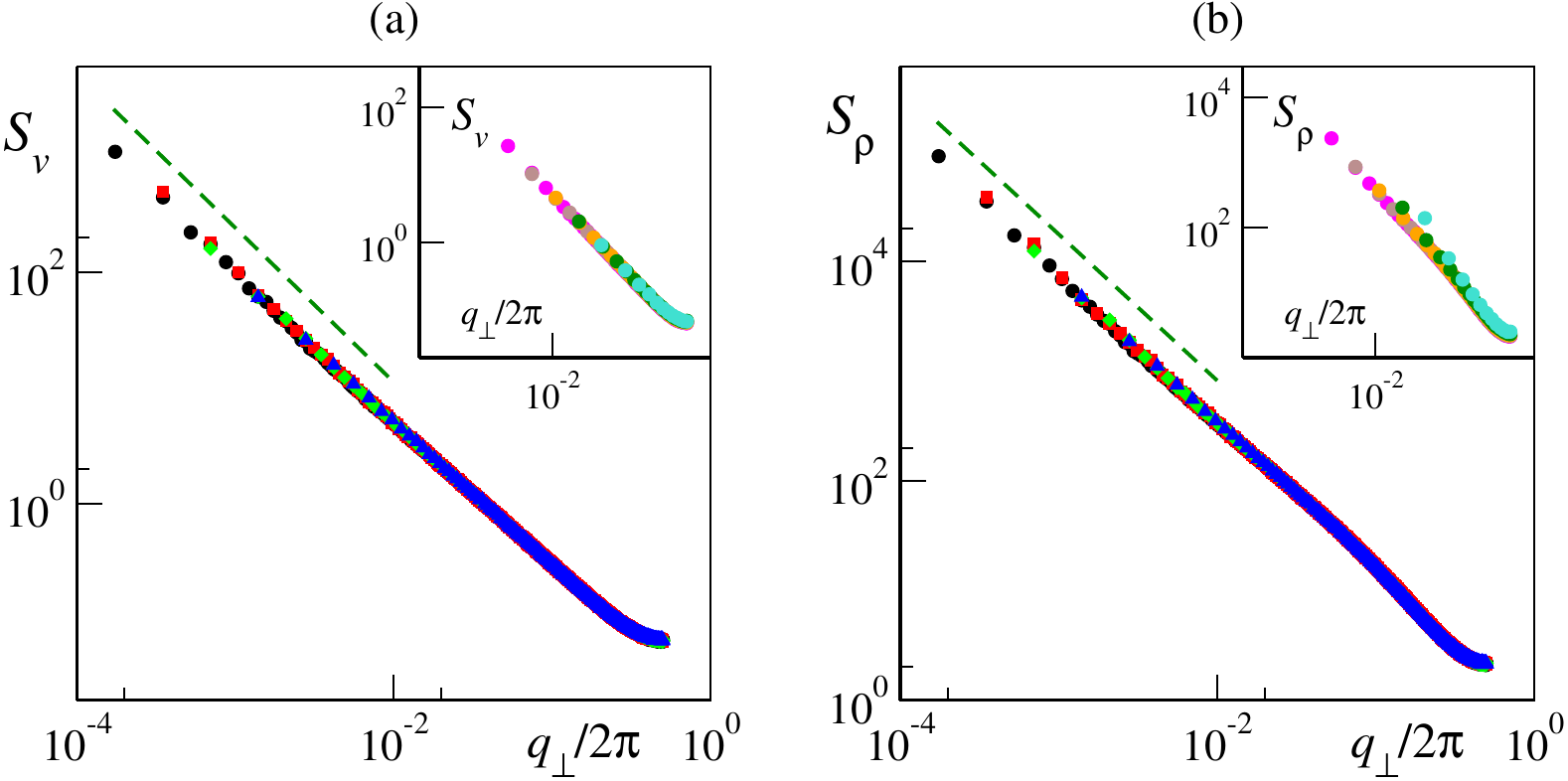}
\caption{{\bf Equal time Fourier correlations in the transverse
  direction.} (a) Transversal velocity correlations and (b) Density
  fluctuation correlations. Transversal system sizes are
  $L_\perp=8192$ (black circles), $L_\perp=4096$ (red squares),
  $L_\perp=2048$ (green diamonds), $L_\perp=1024$ (blue
  traingles). The dashed green line marks a power law
  divergence with exponent $\zeta=1+\xi+2\chi=4/3$ \cite{Solon,
    Benoit2019}. Correlations for smaller transversal separations,
$L_\perp=32, 64, 128,256,512$, are shown in the two insets.}
\label{Fig1}
\end{figure}

In the following, we place ourselves in the polar liquid phase of the
non-confined model \cite{Chate2, VicsekPD} by fixing $v_0=0.5$, $\eta=0.2$ and
$\rho_0=N/(L_\parallel L_\perp)=2$ and we consider more than two
orders of magnitude range of values for the transverse separation
distance, from $L_\perp=32$ to  $L_\perp=8192$, keeping the
longitudinal size constant, $L_\parallel=2048$. For numerical
convenience, we mainly consider uniform
initial positions with alignment parallel to the confining walls, but
we have also verified that also different initial conditions always lead, after a
transient, to a polar state with the mean flocking direction typically
aligned along the longitudinal direction.

In order to measure the density and transversal velocities structure
factors, we first obtain fluctuating fields for the density
$\delta\rho({\bf r})$ and the
perpendicular velocity $v_\perp ({\bf r})$ by coarse-graining the microscopic
particles number and transversal velocities over boxes of unit linear
length. In order to avoid densities inhomogeneities near the confining
walls \cite{Fava2024} we only evaluate the fields in a central channel
of extension $0.7 L_\perp$, thus excluding two regions of size $0.15 L_\perp$
adjacent to the walls. The resulting Fourier space static correlations 
\begin{equation}
S_\rho({\bf q}) \equiv \langle |\delta \hat{\rho} ({\bf
  q})|^2\rangle\;\;\;,\;\;\; S_v({\bf q}) \equiv \langle
| \hat{v}_\perp ({\bf q})|^2\rangle
\end{equation}
are also averaged in time (after a proper transient has been
discarded) over typically $10^6$ timesteps. 

We first fix $q_\parallel=0$ in order to probe correlations in the
transversal direction. Their behavior, reported in Fig.~\ref{Fig1},
does not show any sign of suppression of the small $q_\perp$
divergence which, for sufficiently large separations $L_\perp$, shows an
excellent agreement with the predicted $d=2$ bulk exponent
$\zeta=1+\xi+2\chi=4/3$ \cite{Solon}. This implies that, at least for
typical Vicsek dynamics transversally confined by reflecting walls,
condition (\ref{Cperp}) on the phenomenological parameters
$G_\perp^{(\rho)}$ and $G_\perp^{(v)}$ is
not met.\\

\begin{figure}[t!]
\centering
\includegraphics[width=0.48\textwidth]{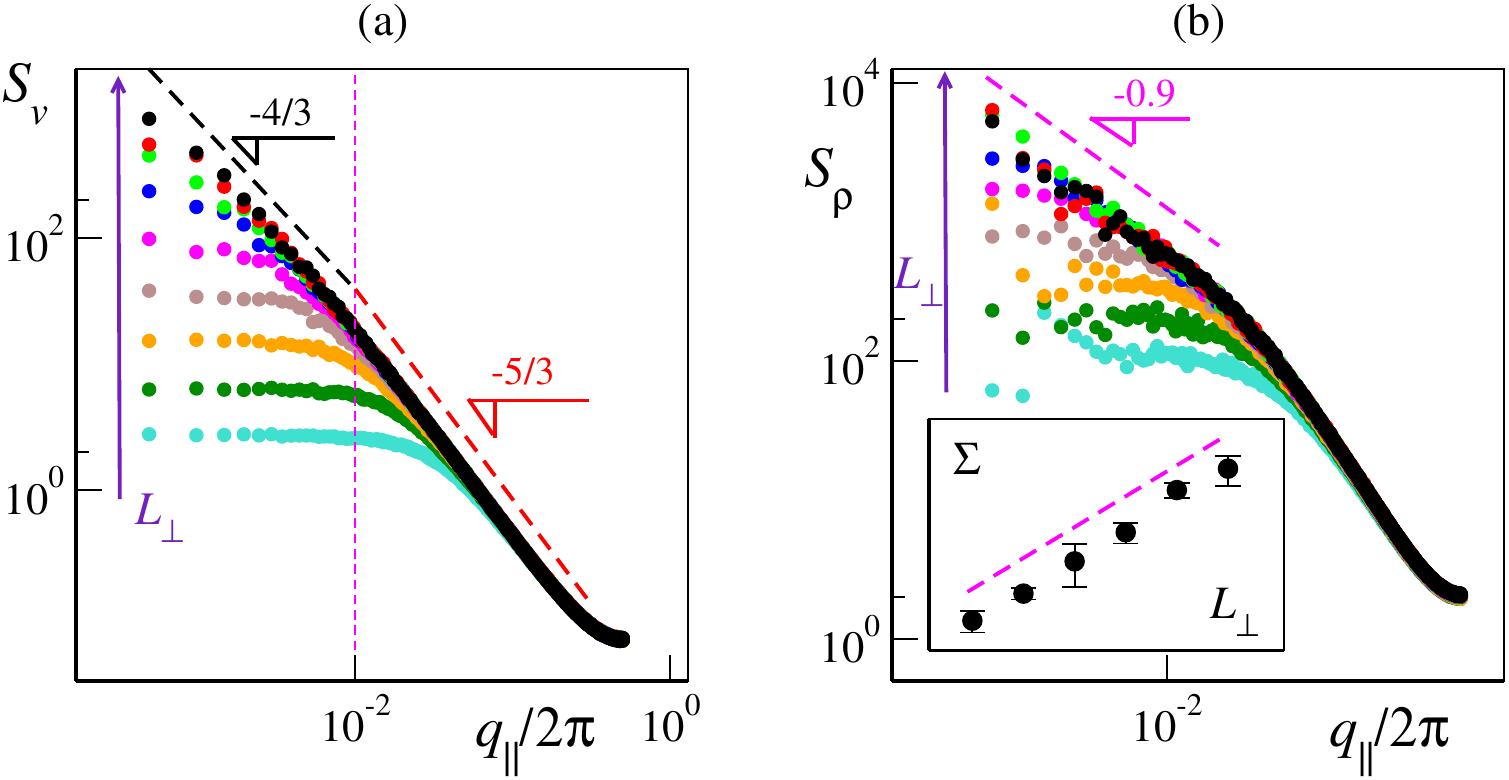}
\caption{{\bf Equal time Fourier correlations in the longitudinal
  direction.}  (a) Transversal velocity correlations. The vertical dashed line marks the crossover scale
  $q_c/(2\pi) = 10^{-2}$ separating the asymptotic regime
  (dashed black line with scaling exponent $4/3$) from the finite size 
  one  (dashed red line with scaling exponent $\approx 5/3$) (see text).
(b) Density
  fluctuation correlations. In both panels transversal separations are, from bottom
  to top, $L_\perp = 32, 64, 128, 256, 512, 1024, 2048, 4096,
  8192$. The dashed magenta line marks an
  algebraic divergence with an exponent $0.9$.
Inset: Scaling of the effective mass term $\Sigma$ as a
  function of transversal separation. The dashed magenta line marks an
  algebraic growth with an exponent $0.9$.}
\label{Fig2}
\end{figure}

The behavior of velocity correlations in the longitudinal direction (see Fig.~\ref{Fig2}a), obtained by setting
$q_\perp=0$, is different, with a clear suppression of the low
$q_\parallel$ divergence for small transversal separations $L_\perp
\lesssim L_\parallel$, where condition
(\ref{Cpar}) is met. Only when $L_\perp
\gtrsim L_\parallel$ one cannot clearly identify a plateau for small
$q_\parallel$, and the structure factor approaches the free theory
algebraic divergence (\ref{Spar}).

We have also tested the density structure factor in the longitudinal
direction fixing($q_\perp=0$), results shown in Fig.~\ref{Fig2}b. Also in this case, a plateau for
$q_\parallel\ll 1$ is evident for $L_\perp
\lesssim L_\parallel$. At larger longitudinal wavenumber, an
anomalous slow scaling behavior $\sim q_\perp^{-\gamma}$, first reported in \cite{Benoit2019},
becomes apparent \footnote{In Ref. \cite{Benoit2019} it was also
  reported an intermediate scaling regime, in qualitative
 agreement with Eq. (\ref{nlnSpar2}) which, however, should be unobservable for $q_\perp \to 0$.}, although our numerical estimates suggest a slightly
larger exponent, $\gamma \approx 0.9$. Applying the scaling arguments
developed in Sec.~\ref{2B} to this empirical scaling behavior
immediately gives the scaling
\begin{equation}
\label{Spar2}
S_\rho(0,q_\parallel,L_\perp^{-1})\sim\dfrac{1}{q_\parallel^{\gamma}+\Sigma(L_\perp)}
\end{equation}
with the anomalous effective mass term scaling
\begin{equation}
\Sigma(L_\perp) \sim L_\perp^{-\gamma}\,.
\label{anomalous}
\end{equation}
When $L_\perp \lesssim L_\parallel$ is possible to measure $\Sigma$
by evaluating $S_\rho(0,q_\parallel, L_\perp^{-1})$ in the
limit $q_\parallel \to 0$. Our results, shown in the inset of
Fig.~\ref{Fig2}b confirm the scaling (\ref{anomalous}) with $\gamma
\approx 0.9$.\\

The scaling behavior predicted by
Eq. (\ref{Spar}) also implies that the velocity structure factor $S_v$
measured for different
transversal separation $L_\perp$ should collapse to a universal curve $f(x)$
when properly rescaled,
\begin{equation}
L_\perp^{-\zeta} S_v(0, L_\perp^{\xi} q_\parallel)\equiv f(x)\,.
\label{collapse}
\end{equation}
When testing this
result by data-collapse, however, one should be aware that the most accurate
numerical simulations of longitudinal structure factors (see
Ref. \cite{Benoit2019}) revealed large finite size effects. In
particular in $d=2$
the anisotropy exponent $\xi$ shows a crossover behavior from a
finite-size (for $q>q_c$) value $\xi \approx 0.8$ to the asymptotic
(for $q<q_c$) one $\xi =1$ as also shown in Fig.~\ref{Fig2}a. For Vicsek dynamics with scalar noise it was
found $q_c/(2 \pi) \approx 10^{-2}$, as marked in Fig.~\ref{Fig2} by the
vertical dashed lines. Transversal correlations, on the other hand, do
not show such crossover and one can confidently use $\zeta=4/3$ over a
wider range of scales. 

\begin{figure}[t!]
\centering
\includegraphics[width=0.48\textwidth]{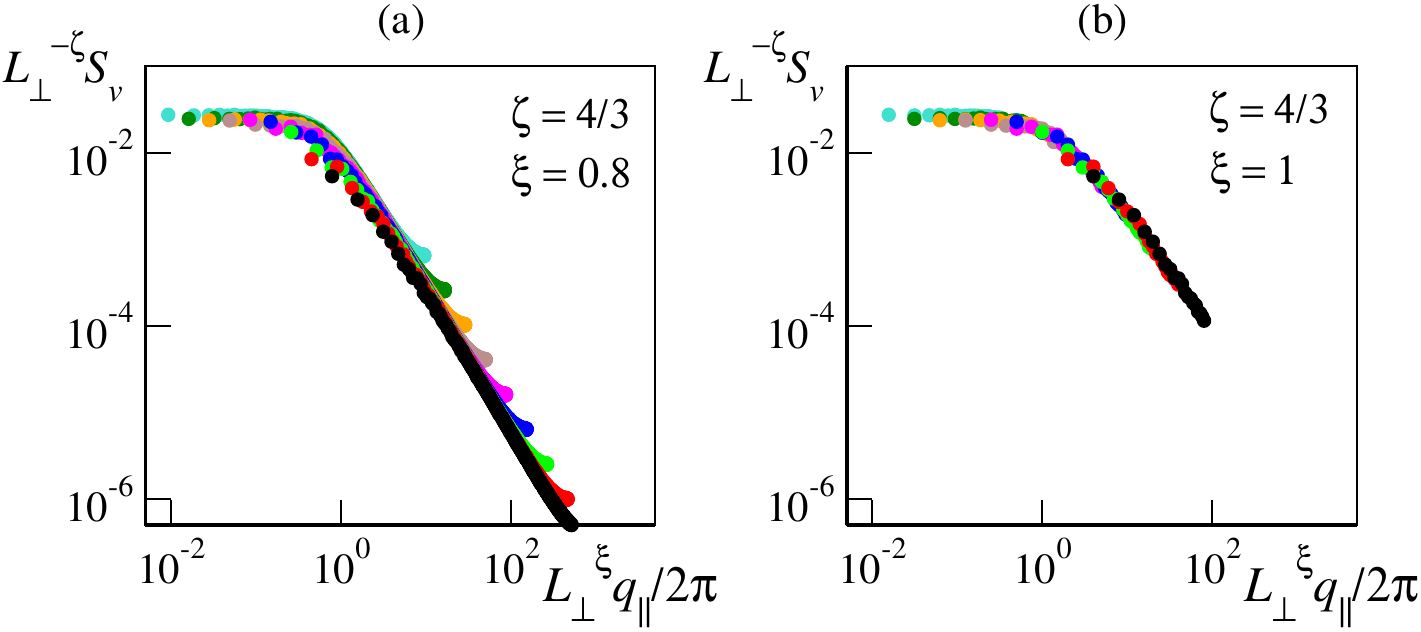}
\caption{{\bf Velocity structure factor data collapses in the
    longitudinal direction.}
  (a) Structure factor data from Fig.~\ref{Fig2} rescaled according to
  Eq. (\ref{collapse}) with scaling exponents $\zeta=4/3$ and
  $\xi=0.8$. (b) Same data, but only for $q_\parallel < q_c$ (see
  text) rescaled with scaling exponents $\zeta=4/3$ and
  $\xi=1$. Color coding for different $L_\perp$ as in Fig. ~\ref{Fig2}a.}
\label{Fig3}
\end{figure}

In Fig.~\ref{Fig3}a we first attempt to rescale the data
of Fig.~\ref{Fig2}a by using the pre-crossover scaling exponents
$\zeta=4/3$ and $\xi=0.8$. While one may observe a reasonable
collapse at large enough $q_\parallel$ values, a
closer inspection reveals a less satisfactory collapse at smaller
wave-numbers. The small $q_\parallel$ collapse may be improved
rescaling the data with the post crossover value $\xi=1$, as shown in
Fig.~\ref{Fig3}b where only the data for $q_\parallel < q_c$ is considered.\\
 
To summarise, numerical simulations of Vicsek dynamics in the confined
polar liquid phase confirm that bulk
correlations in the transversal direction are not affected by the reflecting
boundaries. In the longitudinal direction, our numerics support the scaling form (\ref{Spar}) for the
transversal velocity structure factor and suggest the empirical
anomalous effective mass term (\ref{anomalous}) for density correlations.

\section{Order parameter behavior and equivalence with driven systems}
\label{sec2}

\subsection{Longitudinal response}

We now turn our attention to the behavior of the order parameter in
the presence of transversal confinement. Microscopically, the
instantaneous global
order parameter is defined as
\begin{equation}
\Omega(t)=\frac{1}{N} \sum_{i=1}^N {\bf s}_i^t \,.
\label{SOP}
\end{equation}
The scalar order parameter is given by $\Phi = \langle | \Omega(t) |
\rangle_t$, where $\langle \cdot \rangle$ denotes temporal
averages. Here we are interested in the static longitudinal response, that
is, the difference in the scalar order parameter between the confined
and the bulk system,
\begin{equation}
\delta \Phi(L_\perp) = \Phi(L_\perp)-\Phi(\infty)
\label{response}
\end{equation}
At the hydrodynamic level, the scalar order parameter can be obtained
by the spatial and temporal average of Eq. (\ref{enslave}). Since
linear terms in $\delta \rho$ and ${\bf v}_\perp$ all vanish under
such averages, one is left with 
\begin{equation}
\Phi = p_0 + \langle \delta v_\parallel \rangle \approx p_0 -
\frac{\langle|{\bf v}_\perp|^2\rangle}{2 p_0}\,,
\label{modulo}
\end{equation}
the analogous of the so-called
principle of conservation of the modulus 
that links longitudinal and transversal fluctuations in equilibrium
ferromagnets \cite{Pata}.

\begin{figure}[t!]
\centering
\includegraphics[width=0.435\textwidth]{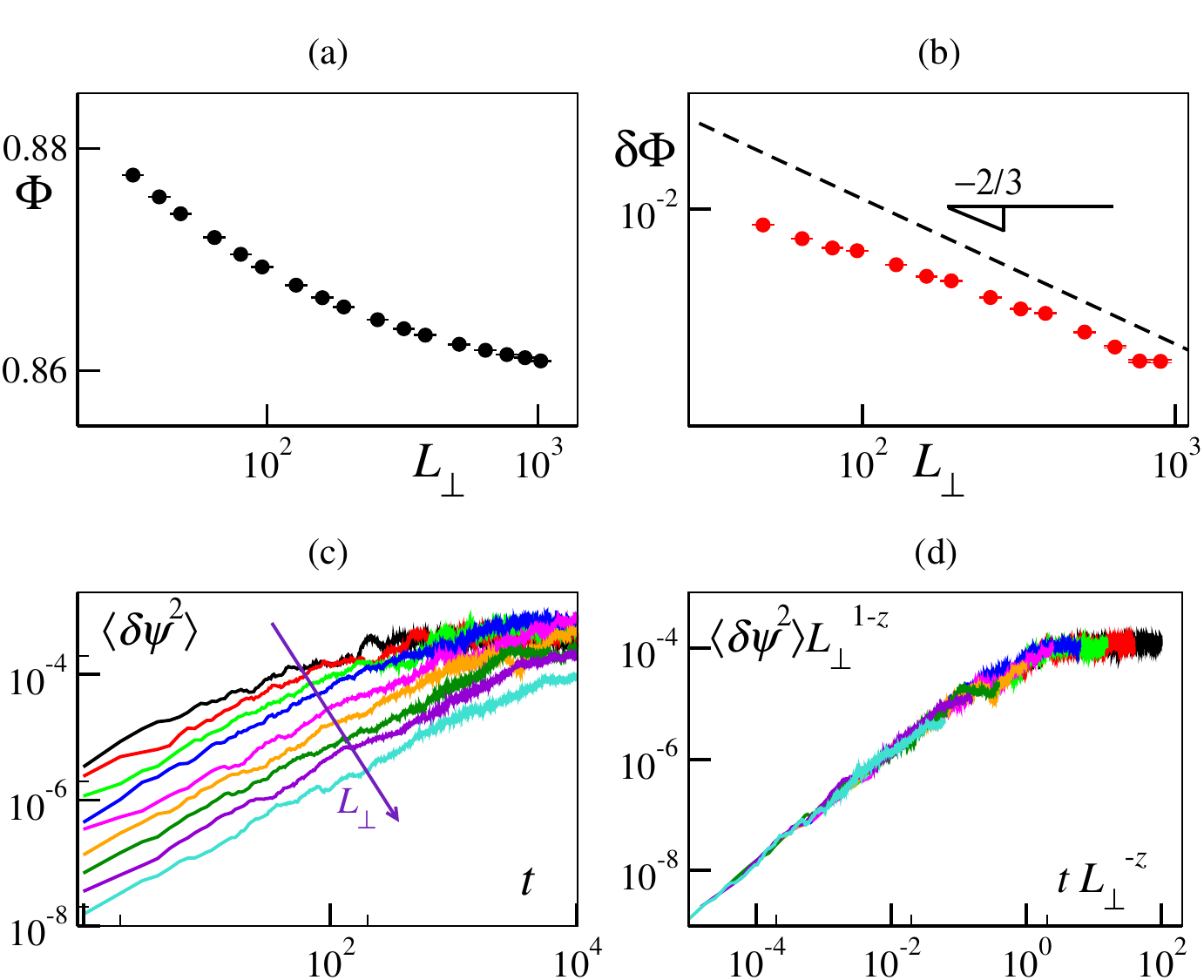}
\vspace{0.25cm}
\caption{{\bf Order parameter scaling.} (a) Scalar order parameter
  $\Phi$ as a function of the transversal separation size. 
(b) Order parameter static response 
$\delta \Phi$, as a function of the transversal separation size,
evaluated from panel (a) by Eq. (\ref{centered}).
System parameters are $L_\parallel=512$,
$\rho_0=2$, $v_0=0.5$ and $\eta=0.21$.
Data has ben averaged over around $10^5$ {\it independent} datapoints.
Error bars mark two standard errors. 
The dashed line marks the expected asymptotic power law decay with an exponent
$2\chi=-2/3$.
(c) Mean squared fluctuations in the flocking direction for
increasing transversal separation sizes, from top to bottom
$L_\perp=32,64,128,256,512,1024,2048,4096, 8192$. (d) Data collapse of the
data in panel (c) according to Eq. (\ref{eq:01}) with dynamical
scaling exponent $z=4/3$. 
Data in panels (c)-(d) has been averaged over $10^2$
different realizations with parameters $L_\parallel=2048$,
$\rho_0=2$, $v_0=0.5$ and $\eta=0.2$.
}
\label{Fig4}
\end{figure}

From Eq. (\ref{response}) it follows that the order parameter response
is thus given by 
\begin{eqnarray}
\delta \Phi(L_\perp) &=& \frac{1}{2 p_0}\left[ \langle |{\bf v}_\perp (\infty)|^2
  \rangle  -\langle  |{\bf
  v}_\perp (L_\perp)|^2\rangle \right] \nonumber\\
& =& \frac{1 }{2 p_0} C_v (0, L_\perp^{-1},{\bm \mu}^{(1)}) 
\end{eqnarray}
where we have recognized the real space correlation function
\begin{equation}
C_v (0, L_\perp^{-1},{\bm \mu}^{(1)}) = \langle|{\bf v}_\perp (\infty)|^2
\rangle - \langle |{\bf
  v}_\perp (L_\perp)|^2\rangle 
\end{equation}
and assumed $L_\parallel \to \infty$. We can deduce the scaling behavior of the response  $\delta \Phi$ by
repeating in real space essentially the same finite size scaling
analysis performed in section \ref{2B}. Performing a DRG step with a
rescaling factor $b$ one obtains
\begin{equation}
C_v (0, L_\perp^{-1},{\bm \mu}^{(1)}) = b^{2\chi} C_v (0, b L_\perp^{-1},{\bm \mu}^{(b)}) 
\end{equation}
where the scaling of the correlation function is determined by the
fact that it just involves a squared field. Choosing as before
$b=L_\perp$ we finally obtain, for a sufficiently large spatial
separation $L_\perp \gg 1$,
\begin{equation}
C_v (0, L_\perp^{-1},{\bm \mu}^{(1)}) = L_\perp^{2\chi} C_v (0, 1,{\bm \mu}^{*}) 
\label{rFP}
\end{equation}
with the r.h.s. parameters now evaluated at the nonlinear fixed
point. Eq.(\ref{rFP}) immediately implies the response
scaling
\begin{equation}
\delta \Phi(L_\perp) \sim L_\perp^{2\chi} \,.
\label{OPscale}
\end{equation}
This asymptotic scaling can be verified by microscopic
simulations of the Vicsek dynamics. Fig.~\ref{Fig4}a shows the
convergence of the scalar
order parameter to its asymptotic value fixed by $p_0$. 
In order to estimate the convergence to this asymptotic value we
consider the centered finite difference 
$\partial_c \Phi(L_\perp)$ of the average order parameter,
which approximates the first derivative of $\Phi(L_\perp)$ up to
corrections of third order in the derivatives. We can then estimate
the response as
\begin{equation}
\delta \Phi \sim L_\perp \partial_c \Phi(L_\perp)\,.
\label{centered}
\end{equation}
From Fig.~\ref{Fig4}b we can see that its asymptotic behavior well
matches our prediction, while a crossover is observed for smaller transversal
separation sizes. This should not come as a surprise: we have seen in
Sec. \ref{2C} that in $d=2$ the anisotropy exponent $\xi$ shows a finite size
crossover from $\approx 0.8$ to its asymptotic value $\xi=1$. Since
$\zeta=d-1+\xi+2\chi=4/3$ remains constant over a wider range of
scales, this implies that also the roughness exponent should
cross-over from the
finite size value $\chi \approx -0.2$ to its asymptotic value $\chi = -1/3$. 

\subsection{Equivalence with  homogeneously driven systems and
  fluctuations in the mean flocking direction}

It is instructive to compare these results with the behavior of
flocking systems in the presence of a homogeneous external field ${\bf
  h}$ of
amplitude $h=|{\bf h}|$ driving each active particle orientation. 
Ref. \cite{Nikos} it was shown that in the linear regime (valid for
small field amplitudes) and large system sizes, the response scales as 
\begin{equation}
\delta \Phi(h) \equiv \Phi(h) - \Phi(0) \sim h^{-2\chi/z}\,.
\label{Phih}
\end{equation}
Similarly, it is known from Ref. \cite{BFG} that an external homogeneous field breaks
explicitely the rotational symmetry and thus induces an
effective mass term $\sim h^{\zeta/z}$ or, thanks to the hyperscaling
relation $\zeta=z$, see Eq. (\ref{scalingexp}), more simply $\sim h$. The
structure factors thus read
\begin{equation}
\label{Hperp}
S({\bf q},h)\sim\dfrac{1}{q_\perp^{\zeta}+D_\perp h^{\zeta/z}}\,,
\end{equation}
in the direction transversal w.r.t. to the external field and
\begin{equation}
\label{Hpar}
S({\bf q},h)\sim\dfrac{1}{q_\parallel^{\zeta/\xi}+D_\parallel h^{\zeta/z}}
\end{equation}
in the direction parallel to ${\bf h}$ (once again, these two scalings
coincide for $\xi=1$).

Comparison between the response scalings (\ref{OPscale})-(\ref{Phih})
and the structure factors (\ref{Sperp})-(\ref{Hperp}) and
(\ref{Spar})-(\ref{Hpar}) immediately suggests the equivalence between
trasversal confinement by reflecting boundaries, a local form of explicit
symmetry breaking, and the presence of a homogeneous external driving field
(which explicitly breaks the rotational symmetry at bulk level) of
amplitude \footnote{Note that this relation holds on dimensional grounds since the field
amplitude scales as the inverse of a time \cite{Nikos}.}
\begin{equation}
h \sim L_\perp^{-z}\,.
\label{equivalence}
\end{equation}

We test this equivalence on the fluctuations of the fluctuations of
the mean flocking direction $\psi(t) = \mbox{Arg}[\Omega (t)]$. In Ref. \cite{BFG}, it has been
shown by mean-field like approximations of the microscopic VM dynamics, that $\psi(t)$ obeys an
Ornstein-Uhlenbeck process. In particular, we are interested in the
fluctuations of the mean flocking direction, $\delta \psi(t) =
\psi(t)-\psi(0)$ with the initial condition $\psi(0)$ aligned in the
external field direction. The mean squared fluctuations are thus given by
\begin{equation}
\langle \delta \psi(t)^2 \rangle \approx 
\frac{D_\psi}{h'}\left(1-e^{-2 h' t} \right) 
\label{eq99}
\end{equation}
where $h' = h/\bar{m}$, with $\bar{m}$ being the average number of interacting
particles in the Vicsek dynamics (\ref{VM}), plays the role of the
confining potential stiffness \footnote{Note that this potential
  confines the mean flocking orientation, not the position of the
  active particles.} and 
\begin{equation}
D_\psi=\frac{\eta^2 \pi^2}{6 N}
\end{equation}
is (twice) the effective
diffusion acting on the mean flocking orientation. Eq. (\ref{eq99}) implies that $\langle \delta \psi^2 \rangle$ will grow
linearly as $2D_\psi t$ on short timescales, $t \ll 1/(2h')$, eventually saturating to a constant
value $D_\psi/h'$ for $t \gg 1/(2h')$. In practice, the short time
dynamics of $\psi(t)$ is diffusive and cannot be distinguished from
the zero field behavior, while at large enough time the external field
suppresses orientation diffusion and acts as a confining potential on
the mean flocking direction.

We now consider fluctuations of the mean flocking directions in the
absence of an external field but for a transversally confined flock. Carrying on the correspondence expressed by Eq. (\ref{equivalence})
and using $N=\rho_0 L_\parallel L_\perp$ one gets by direct
substitution in Eq. (\ref{eq99}) 

\begin{equation}
\langle\delta\psi(t)^2\rangle=\dfrac{\eta^2 \pi^2}{6\rho_0
  L_\perp
  L_\parallel}\dfrac{\bar{m}}{L_\perp^{-z}}\left[1-\mbox{exp}\left(-\frac{2
      L_\perp^{-z} } {\bar{m}} t \right)\right] 
\label{eq:00}
\end{equation}
or, introducing the scaling function
\begin{equation}
f(x)\equiv\dfrac{\eta^2 \pi^2 \bar{m}}{6\rho_0 
  L_\parallel}\left[1-\mbox{exp}\left(-\frac{2}{\bar{m}} x \right)\right] \,,
\end{equation}
\begin{equation}
\langle\delta\psi(t)^2\rangle=L_\perp^{z-1}f( L_\perp^{-z} \,t)\,.
\label{eq:01}
\end{equation}
Transversal confinement between parallel reflecting walls thus
suppresses diffusion of the mean flocking directions at large enough
times, with
\begin{equation}
\langle\delta\psi(\infty)^2\rangle \sim L_\perp^{z-1}
\label{eq:02}
\end{equation}
while at short times ($t \ll L_\perp^z \bar{m}/2$) one has a diffusive behavior with
\begin{equation}
\langle\delta\psi(t)^2\rangle \approx \dfrac{\eta^2 \pi^2}{3\rho_0
  L_\parallel L_\perp}\,t\,.
\end{equation}
\label{eq:03}
Microscopic numerical simulations, reported in Fig.~\ref{Fig4}c,
clearly show these two regimes. In particular, data collapse (see Fig.~\ref{Fig4}d) confirms the scaling form (\ref{eq:01}), thus
supporting the equivalence conjectured in (\ref{equivalence}).

\section{Discussion}
\label{sec4}
In this work, we have shown by analytical and numerical arguments that
transversal confinement between parallel reflecting walls separated by
a distance $L_\perp$ introduces an
effective mass term ${M_c} \sim L_\perp^{-\zeta}$ damping the free theory
Nambu-Goldstone mode. 

The effect of confinement on bulk connected correlations can be deduced by
standard finite size scaling analysis under the rather generic
assumption that the transversal boundaries break the rotational symmetry
of the active particles self propulsion orientation. In principle the mass
term should suppress the small wavenumber divergence of fluctuations correlations in
Fourier space (or, equivalently, introduces an exponential cut-off in the
real space connected correlations), introducing a crossover towards a
finite value,
\begin{equation}
S({\bf q} \to 0) \sim M_c^{-1}\,.
\end{equation}
This crossover is however controlled by phenomenological constants
which in principle depends on microscopic parameters and on the details
of the interaction with the confining boundaries. In the transversal
direction, where wave-numbers are bounded from below due to the
finiteness of the sytem, this implies that the saturation regime does
not appear for a phenomenological constants $G_\perp$ of order one or
smaller. We have verified numerically that this is for instance the
case for the standard VM model confined between parallel reflecting
walls. \\
In the longitudinal direction, on the other hand, such a crossover can be
observed in a narrow enough ring configuration, that is for $L_\perp
\lesssim L_\parallel$. Density correlation in the longitudinal
direction, however, seem to be 
characterized by an anomalous effective mass term $\Sigma \sim
L_\perp^{-\gamma}$, with $\gamma \approx 0.9$ an empirical scaling
factor of unknown origin.\\

Confinement also increases the scalar order parameter values, which
shows an algebraic decay to its asymptotic value with increasing
separation sizes,
\begin{equation}
\Phi(L_\perp) - \Phi(L_\perp \to \infty) \sim L^{2\chi}\,.
\label{FSOP}
\end{equation} 
Since this result has been obtained only by finite scaling analysis,
it does not depend in any way on the nature of boundary conditions,
and should also apply to the standard numerical setup of finite
systems with periodic boundary conditions (PBC), as it has been
already realized in \cite{Duan}. Indeed, numerical
simulation of the VM in two dimensional tori of linear size $L$ show a power
law decay of the scalar order parameter to its asymptotic value, $\Phi(L) - \Phi(L \to \infty) \sim L^{\alpha}$, with
an exponent $\alpha \approx -0.64$ \cite{Chate2020} compatible with
the theoretical prediction $\alpha=2\chi=-2/3$. Interestingly, this
result suggest an alternative and more robust way to estimate
numerically the theory scaling exponent from the finite size scaling
of global observables rather than from the measure of the small
wavelength behavior of correlation functions.

Altogether, these results suggest an equivalence between the effects
of an explicit symmetry breaking due to boundary conditions and the
one induced by an homogeneous (and small) external driving of
amplitude $h$. Both induce an effective mass term, with the
equivalence $M_c \sim h \sim L_\perp^{-z}$. 
This mass term, finally, constrains the mean flocking
direction $\psi(t)$ fluctuations acting as an effective harmonic potential
stiffness which, at large times, prevents the diffusion of $\psi(t)$
that characterize free finite flocks. 

We believe our results could be of experimental relevance and can be tested by confining 
flocking systems such as active colloids \cite{Bartolo2013} or flocking epithelial
tissues \cite{Giavazzi}.

\begin{acknowledgments}
We
thank B. Mahault for
valuable discussions. We acknowledge support from PRIN
2020PFCXPEAG.
\end{acknowledgments}


\bibliography{biblio_abbr}

\end{document}